
\input harvmac
\def\s{\sqrt}
\def\na{\nabla}
\def\({\left (}
\def\){\right )}
\gdef\journal#1, #2, #3, 19#4#5{
{\sl #1 }{\bf #2}, #3 (19#4#5)}

\lref\btz{M. Banados, C. Teitelboim, and J. Zanelli,
\journal Phys. Rev. Lett.,
69, 1849, 1992.}

\lref\btzprd{M. Banados, M. Henneaux, C. Teitleboim, and J. Zanelli,
\journal Phys. Rev., D48, 1506, 1993.}

\lref\carlip{For a nice review see S. Carlip, hep-th 9506079.}

\lref\witten{E. Witten, \journal Phys Rev., D44, 314, 1991.}

\lref\hw{G. Horowitz and D. Welch, \journal Phys. Rev. Lett., 71, 316,
1993.}

\lref\chan{K. Chan, gr-qc 9509032.}

\lref\todd{See for example, Hugget and Todd, {\it Introduction to Twistor
Theory}, second edition, Cambridge University Press, 1994.}

\lref\kk{M. Kamata and T. Koikawa, \journal Phys. Lett., B353, 196,
1995.}

\lref\jhgh{See for example,
J. Horne and G. Horowitz, \journal Nucl. Phys., B368, 444,
1992.}

\lref\brown{J.D. Brown, Ph.D. thesis, University of Texas at Austin
1985; published as a book {\it Lower Dimensional Gravity} (World
Scientific, Singapore, 1988).}

\lref\djh{S. Deser, R. Jackiw, and G 't Hooft, \journal Ann. Phys.,
152, 220,1984. }

\lref\clement{G. Clement, gr-qc 9510025.}

\Title{\vbox{\baselineskip12pt\hbox{UCSBTH-95-33}
\hbox{hep-th/9510181}}}
{\vbox{
\centerline{Magnetic Solutions to 2+1 Gravity}}}

\centerline{Eric W. Hirschmann}
\centerline{\sl Department of Physics}
\centerline{\sl University of California}
\centerline{\sl Santa Barbara, CA 93106-9530}
\centerline{\sl ehirsch@dolphin.physics.ucsb.edu}
\tenrm

\bigskip

\centerline{Dean L. Welch}
\centerline{\sl Department of Physics}
\centerline{\sl University of California}
\centerline{\sl Santa Barbara, CA 93106-9530}
\centerline{\sl dean@cosmic.physics.ucsb.edu}

\bigskip
\centerline{\bf Abstract}
We report on a new solution to the Einstein-Maxwell equations in
2+1 dimensions with a negative cosmological constant.
The solution
is static, rotationally symmetric and has a non-zero magnetic field.
The solution can be interpreted as
a monopole with an everywhere finite energy density.

\newsec{Introduction}
Studying physics in spacetimes with dimensions less than four has often
proved
useful. While the study of such systems is of intrinsic interest,
one usually has the hope that the properties of these lower
dimensional systems will mimic the properties of some corresponding
four-dimensional system.
Witten's discovery \witten\ of a black hole solution in two-dimensional
string theory has
sparked renewed interest in lower dimensional gravity. This
solution has been used to study problems which have been intractable in
four dimensions such as black hole
information loss.

Another lower-dimensional black hole solution that has generated a great
deal of interest is the three-dimensional black hole discovered by
Ba\~nados, Teitelboim and Zanelli (BTZ) \btz\ \btzprd . This spacetime
is a solution to Einstein gravity with a negative cosmological constant,
it is also known that this solution can be formulated as a string theory
solution \hw .
Like the two dimensional black hole,
this solution has been studied with the hope of shedding light on
problems in four-dimensional gravity. This hope is supported by the fact
that there
are striking similarities
between some of these recently discovered three-dimensional
solutions and their four
dimensional counterparts.
However, despite these similarities one should bear in mind that,
there are some important
differences, not the least of which is that the universe we live in is
four dimensional.
Nevertheless,
in the more simplified realm of three dimensions
we might reasonably hope to obtain some insight into the
nature of gravity and quantum gravity in particular \carlip .

The discovery of the BTZ black hole has spawned efforts to find other
solutions to the three-dimensional Einstein equations as well as solutions
to various
generalizations of them coupled to a variety of matter fields.  One such
solution is the static electrically charged black hole originally
discussed by BTZ \btz .
This solution is specified by three parameters, a mass parameter $M$, a
charge $Q$, and a ``radial parameter" $r_0$.
To see that one needs this radial
parameter it is sufficient to observe that while
the energy density in the
electromagnetic field approaches zero asymptotically, the rate at which
it does is sufficiently slow so that the total energy in the
electromagnetic field outside of any finite radius diverges.
This is most easily seen by observing the behavior of the quasilocal
mass as a function of $r,$ $M_{ql}=M+Q^2\ln (r/r_0)$
The parameter
$r_0$ serves to determine how much of the energy in the electromagnetic
field is included in the mass parameter $M$.\footnote{$^{1}$}{In other
words, for a given solution, changes in $M$ can be compensated for by
changes in $r_0$, see \btz .}
Depending on the
values of the parameters the charged BTZ solutions may have two, one,
or no horizons. It is natural to identify a charged
solution with a single horizon as an
extremal black hole.
This identification is supported by the observation that such a
solution has zero Hawking
temperature, as does the extremally charged Reissner-Nordstrom black
hole.

Although the static, electrically charged
solution is similar in many ways to the Reissner-Nordstrom black hole in
four dimensions, there are some important differences. The most
obvious difference is that the three-dimensional black hole is
asymptotically anti-de Sitter space while the Reissner-Nordstrom
solution is asymptotically flat. Another difference that we have just
seen is that the static solution has a quasilocal mass that
diverges at infinity whereas the quasilocal mass of the
four-dimensional charged black hole approaches a constant
asymptotically \todd\ .  An additional difference that is the
consideration of
this work is that the Reissner-Nordstrom black hole
can have electric or magnetic charge, as well as both.  Due to the
invariance of the Maxwell equations under a duality transformation, the
form of the metric for a Reissner-Nordstrom black hole is the same for
an electrically charged solution and a magnetically charged solution.
The reason is that
in four dimensions both the Maxwell tensor and its dual
are two-forms.
However, no such transformation exists for the Maxwell equations in
three dimensions because the Maxwell tensor is a two-form and its dual
is a one-form.  One is naturally led to ask whether the solutions
to the Einstein-Maxwell equations in three dimensions are different if
one assumes that they possess a magnetic as opposed to electric charge.
We examine this question in this paper and report that the solutions
are quite different.  Whereas the electric solution may be a black
hole provided the charge is not too large, the magnetic solution that we
present is not black hole for any value of the magnetic charge.
This magnetic solution is both static and rotationally
symmetric. In addition, it has finite
energy density.  We interpret it as a magnetic monopole.\footnote{$^2$}
{We are using the term monopole a bit loosely here.  The solution is
certainly magnetic and particle-like, but the fact that we are in two
spatial dimensions suggests that the solution is perhaps a bit more reminiscent
of a Neilson-Oleson vortex solution.}

Parenthetically, we remark that there has been some recent discussion of
stationary generalizations of the electrically charged solution \kk\
\clement .
For a rotating and charged solution, one would
expect there to be both an electric and a magnetic field.
Indeed, these new rotating solutions would appear to possess both.
However, it was incorrectly reported in
\kk\ that their extremal solution has a finite angular momentum.  Chan
\chan\ showed that the angular momentum of the solution in \kk\
actually diverges logarithmically at infinity. Given that the
mass of a static electrically charged solution also diverges
logrithmically at infinity, we believe that
the divergence in the angular momentum
is not physically unreasonable.

\newsec{The Einstein Equations and Their Solution}

We begin with the action for Einstein gravity coupled to a $U(1)$
gauge field with a negative cosmological constant.
\eqn\action{
S = {1 \over 4}\int d^{3}x \s{-g} (R - 2\Lambda - F^2) }
where $\Lambda = -1/l^2$ is the negative cosmological constant,
$F$ is a two-form and we have set Newton's constant to be $1/4\pi$.
The equations of motion derived from the action
are the Einstein equations
\eqn\eineqn{
R_{ab} - {1 \over 2}g_{ab}R + \Lambda g_{ab} = 2 T_{ab}
}
and the Maxwell equations
\eqn\max{
\na_{a} F^{ab} = 0
}
with the stress tensor of the electromagnetic field given by
\eqn\stress{
T_{ab} = F_{ac} F_{bd} g^{cd} - {1 \over 4} g_{ab} F^2  .
}

We assume the spacetime is both static and rotationally symmetric,
implying the existence of a timelike Killing vector and a spacelike
Killing vector. In the coordinate basis we use, these vectors will be
$\partial/\partial t$ and $\partial/\partial\phi$ respectively.
These symmetries allow us to
write our metric in the following form \kk\
\eqn\met{
ds^{2} = -N(r)^{2} dt^{2} + L(r)^{-2} dr^{2} + K(r)^{2} d\phi^2 .
 }
Using this metric and the substitutions $E_r = {L\over N}F_{t r}$ and
$B={L\over K}F_{ r  \phi},$ ($E_r$ and $B$ are the components of the Maxwell
tensor measured in an orthonormal basis)
the Einstein-Maxwell equations, \eineqn\ through
\stress\ become
\eqna\comps
$$\eqalignno{
R_{tt} & =N^2  L^2\left( {N'K' \over NK} + {L'K' \over
     LK} + {N'' \over N}\right)   \cr
  & = N^2\( {2 \over l^2} + 2 B^2 \)  &\comps a \cr
R_{\phi \phi} & =  -K^2 L^2\left({N'K' \over NK} + {L'K'
\over LK} + {K'' \over K}\right) \cr
  & = K^2\( -{2 \over l^2} + 2 E_r^2\) &\comps b \cr
R_{t\phi } & =  0 \cr
  & =  -2 E_r B &\comps c \cr
G_{rr} & = {N' K' \over N K} \cr
  & =  {1 \over L^2} \( {1 \over l^2} + B^2 - E_r^2 \)  &\comps d \cr
}$$
where the prime indicates differentiation with respect to $r$.
In addition, we can write the Maxwell equation as
\eqn\maxwellnew{
\partial_a (\s{-g} g^{ab} g^{cd} F_{bc}) = 0
}
which, upon integration
yields
\eqn\fields{
E_r = {C_1 \over K},
\qquad  B = {C_2 \over N} .
}

We have made no assumptions other than the fact that
the spacetime is static and rotationally symmetric. The $R_{t
\phi }$ equation implies that one or both of the electric and
magnetic fields must be zero.\footnote{$^{3}$}{This will no
longer be necessarily
true for the rotating case.}  The
electric case has previously been discussed.
However, we are interesed in magnetically charged
solutions, so we make the assumption that $C_2\neq 0$
which immediately implies that $C_1=0$.

Using our form for $B$, we can solve our equations as follows.
We make the substitution
\eqn\subst{
L(r) = KN f(r)
}
where $f(r)$ is a function which can be freely specified.
We can combine
\comps a\ and \comps d\ to get an equation in $N$ and $f$
\eqn\cleanup{
{f \over 2} (N^2)' = a_0
}
where $a_0$ is an integration constant.  Likewise, equation
\comps d\ will now yield
\eqn\cleanupp{
{f \over 2} (K^2)' = {1 \over l^2} + {2 C_{2}^{2} \over N^2}  .
}
For the simple choice $f(r) = 1/r$, the metric coefficients become
\eqn\coef{\eqalign{
N^2(r)  = a_0 r^2 + a_1  \qquad
K^2(r)  = {1 \over l^2 a_0} r^2 + {C_{2}^{2} \over a_{0}^{2}} \ln |a_0 r^2
    + a_1|  + a_2
} }
This solution is asymptotic to anti-de Sitter space with curvature
$-1/l^2.$ The integration constant $a_2$ can be absorbed into the other
integration constants together with a rescaling of $r^2$,
so we will choose it to be zero.
We choose a normalization of $t$ so that as $r$ becomes large, $g_{tt}$
approaches $-r^2/l^2$.  This is equivalent to choosing $a_0=1/l^2.$
When the magnetic field is zero, $C_2=0,$ this solution is a
three-dimensional black hole with mass equal to $-a_1.$ Therefore,
we set $a_1=-M.$
Asymptotically,
the Maxwell field looks like that of a magnetic point charge, so we
set $C_2 = Q_{m}/l^2$ ($Q_m$ representing the magnetic
charge).
The metric \met\ is now in the form
\eqn\newmet{\eqalign{
ds^2 &= -(r^2/l^2 - M) dt^2 + r^2 (r^2/l^2 - M)^{-1} (r^2 +
    Q_m^2 \ln |r^2/l^2 - M|)^{-1} dr^2 \cr
    &\qquad+ (r^2 +
    Q_m^2 \ln |r^2/l^2 - M|) d\phi^2  .
}}
For future
convenience, we make the definition,
$r_{+}^{2} = M l^2$.
For $Q_{m}=0$, the metric \newmet\
is identical to the nonrotating
BTZ solution, as we would expect.  However,
the presence of a nonzero magnetic charge drastically changes the spacetime.

The nonrotating three-dimensional black hole obtained by setting
the magnetic charge to zero has an event horizon at
$r = r_+$.  However,
there is no event horizon for the case of nonzero magnetic
charge. In particular,
we do not have a magnetically charged three-dimensional black
hole. This can be seen as follows.  The
$g_{\phi\phi} = K^2$ term becomes zero for some value of $r$ which
we call $ \bar{r}.$ By definition $\bar r $ satisfies
\eqn\rbar{
\bar{r}^2 + Q_m^2\ln |{\bar{r}^2 -r_{+}^{2} \over l^2}| = 0.
}
Clearly $\bar{r}$ is constrained to be between $r_+$ and
$\s {r_{+}^2 + l^2}$.
Not only does $g_{\phi \phi}$ change sign as $r$ becomes less
than $\bar{r},$ but $g_{rr}$ changes sign as well. One can see
that naively
using these coordinates for $r<\bar{r}$ leads to an apparent
signature change. This shows that we must choose a different
continuation for $r\le \bar{r}$ \jhgh .
We now introduce a new set of coordinates that will
show the spacetime is complete
for $r\ge \bar{r}.$
A ``good" set
of coordinates which allows us to cover our spacetime is found by
letting
$$
x^2 = r^2 -\bar{r}^2
$$
The metric with this new coordinate then becomes
\eqn\newmetric{\eqalign{
ds^2 =&-{1\over l^2}(x^2 + \bar{r}^2 - r_{+}^{2}) dt^2
    + (x^2 + Q_m^2 \ln [1 + {x^2 \over \bar{r}^2 - r_{+}^2}]) d\phi^2 \cr
&+ l^2 x^2 (x^2 + \bar{r}^2 - r_{+}^{2})^{-1}
     (x^2 + Q_m^2 \ln [1 + {x^2 \over \bar{r}^2 - r_{+}^2}]
     )^{-1} dx^2
}}
where our coordinate $x$ ranges between zero and infinity.
This coordinate system now covers the complete spacetime.
Timelike geodesics can reach the
origin, $x=0,$ in a finite proper time and null geodesics can reach the origin
in a finite affine parameter. The components of the Ricci tensor
measured in an orthonormal basis that is parallel propagated along a
timelike geodesic are well behaved everywhere.
In three dimensions the curvature
is completely determined by the Ricci tensor, so
the fact that the Ricci tensor is well behaved shows that this
spacetime has no curvature singularities. Similarly, the components of
the electromagnetic field strength are well behaved in this basis.

However, one can see that at $x=0$, we will have a conical singularity
unless we identify the coordinate $\phi$ with a certain period.  The
period is found to be
\eqn\period{
T_{\phi} = 2\pi {e^{\beta/2} \over 1 + Q_m^2 e^{\beta}/l^2}
}
where $\beta = \bar{r}^2 / Q_m^2$.  The strange thing about
this period reveals itself when we examine its behavior for limiting
values of $Q_{m}$.
As $Q_m $ approaches infinity the
period becomes zero.
This is what one might expect because
this is the limit in which the magnetic charge is approaching infinity.
However, as $Q_m $ approaches zero the same thing happens, the period
of the coordinate $\phi $ approaches zero again.
This is very surprising since the $Q_m =0$ solution is a
three-dimensional black hole (with no magnetic charge)
and this looks nothing like the $Q_m \ne
0$ solution in the limit as $Q_m $ approaches zero.  While it often
occurs that ``the limit of a theory is not the theory of the limit,"
in the case considered here the difference is quite striking.

\newsec{The Spacetime for Negative $M$}

In the previous analysis we have been assuming that $M\ge 0.$ We now
briefly consider magnetic solutions with negative $M$.  As observed in
\btz , the BTZ solution for $-1<M<0$ reduces to a solution with a naked
conical singularity. Such solutions were studied in Refs. \djh\
\brown .
For $M$ in this range the magnetic solution, \newmet\ with $Q_m\ne 0,$
is continued in the same way as for the case of nonnegative $M.$
However, now we have $r^2+l^2|M|$ appearing in the logarithm in
\newmet , so ${\bar r}$ is constrained to be between $0$ and
$l\sqrt {1-|M|}$.

For $M=-1$ the BTZ solution is anti-de Sitter space. It is interesting
to observe that the magnetic solution with $M=-1$ is already complete
with no apparent signature change in the metric.  Equivalently, the
analysis in the previous section applies, but with ${\bar r}=0.$
In particular, note that the period of $\phi$ needed to avoid a
conical singularity is
\eqn\perioda{T_{\phi }(M=-1)={2\pi \over 1+Q_m^2/l^2}}
for $M=-1.$ This period still approaches zero as $Q_m$ approaches
infinity, but as $Q_m$ approaches zero \perioda\ goes to a constant.
This behavior of the period of $\phi$ is what intuition tells
one it should be (one should bear in mind that the coordinate $\phi $
is not identified for anti-de Sitter space).

Finally, consider the magnetic solution with $M<-1.$ This space is
incomplete if one only considers non-negative values of $r^2.$
To complete this space we must allow $r^2$ to become negative. This
is not as strange a thing to do as it may seem (in particular it
does not require us to consider complex coordinates).  Note that $r$ in
\newmet\ only appears as $r^2$, this indicates that $r^2$ may be a more
natural radial coordinate (see also the transformations in \hw\ ).
The coordinate transformation $r^2=x^2-l^2|M|$ leads to the desired
result. If we allow $x$ to range over all non-negative values the space
will be complete. The remainder of the analysis carries through as with the
$M=-1$ case.

\newsec{Conclusions}
In this paper we have presented a new solution to the Einstein-Maxwell
equations in 2+1 dimensions in the presence of a negative cosmological
constant.  It is static, rotationally symmetric and magnetically
charged.  The nature of the spacetime is very different from that for
the electrically charged BTZ solution.  In the latter case, for a nonzero
region of parameter space the solution is a
black hole.  In contrast, the magnetic
case has no event horizon and is particle-like.

There are several other things one might like to know about this
solution.  One possibility would be to understand the motion of
magnetically charged particles in
this spacetime.  Another interesting question would be whether
this solution could be generalized to one that included rotation.

\bigskip

\centerline{Acknowledgments}
We wish to thank Gary Horowitz for helpful discussions and Doug Eardley
for reading a preliminary draft.  This work was
supported by
NSF Grant No. PHY-9008502.
\vskip .5cm

\listrefs

\end